\documentclass[aps,10pt,a4paper,showpacs,pre,twocolumn,            superscriptaddress]{revtex4-1} 

\usepackage{chemformula} % Formula subscripts using \ch{}
\usepackage[T1]{fontenc} % Use modern font encodings
\usepackage{amssymb}
\usepackage{mathrsfs}
\usepackage{stmaryrd}
\usepackage{amsmath}
\usepackage{subfigure}

\usepackage{graphicx}
\usepackage{color}

\usepackage{enumerate}
\usepackage{fancyhdr}

\newcommand{\mr}{\mathrm}
\newcommand{\eb}{\varepsilon_{\mr{sp}}}
\newcommand{\csa}{c_{\text{$s_A$}}}
\newcommand{\csb}{c_{\text{$s_B$}}}

\definecolor{amber}{rgb}{1.0, 0.49, 0.0}

\newcommand{\fs}[1]{\textcolor{black}{#1}}
\newcommand{\xc}[1]{\textcolor{black}{#1}}
\newcommand{\pb}[1]{\textcolor{black}{#1}}

\newcommand{\rem}[1]{}

\newcommand{\xcrev}[1]{\textcolor{black}{#1}}
\newcommand{\fsrev}[1]{\textcolor{black}{#1}}

\begin{document}

\author{Xinxiang Chen}
\email{xichen@uni-mainz.de}
\affiliation{Institute of Physics, Johannes Gutenberg-University, 55099 Mainz, Germany}

\author{Jude Ann Vishnu}
\affiliation{Institute of Physics, Johannes Gutenberg-University, 55099 Mainz, Germany}

\author{Pol Besenius}
% \email{besenius@uni-mainz.de}
\affiliation{Department of Chemistry, Johannes Gutenberg-University, 55099 Mainz, Germany}

\author{Julian K\"onig}
% \email{Julian.koenig@uni-wuerzburg.de}
\affiliation{Institute of Molecular Biology, 55128 Mainz, Germany}
\affiliation{Theodor Boveri Institute, Biocenter, University of W\"urzburg, 97074 W\"urzburg, Germany}
%\affiliation{Theodor Boveri Institute, Biocenter, University of W\"urzburg, 97074 W\"urzburg, Germany}

\author{Friederike Schmid}
\email{friederike.schmid@uni-mainz.de}
\affiliation{Institute of Physics, Johannes Gutenberg-University, 55099 Mainz, Germany}

%%%%%%%%%%%%%%%%%%%%%%%%%%%%%%%%%%%%%%%%%%%%%%%%%%%%%%%%%%%%%%%%%%%%%
%% The document title should be given as usual. Some journals require
%% a running title from the author: this should be supplied as an
%% optional argument to \title.
%%%%%%%%%%%%%%%%%%%%%%%%%%%%%%%%%%%%%%%%%%%%%%%%%%%%%%%%%%%%%%%%%%%%%
\title[An \textsf{achemso} demo]
  {%Specific binding effect causes percolation without phase separation in RNA-protein mixtures
Sol-gel transition in heteroassociative RNA-protein solutions:
A quantitative comparison of coarse-grained simulations and the
Semenov-Rubinstein theory}

%%%%%%%%%%%%%%%%%%%%%%%%%%%%%%%%%%%%%%%%%%%%%%%%%%%%%%%%%%%%%%%%%%%%%
%% Some journals require a list of abbreviations or keywords to be
%% supplied. These should be set up here, and will be printed after
%% the title and author information, if needed.
%%%%%%%%%%%%%%%%%%%%%%%%%%%%%%%%%%%%%%%%%%%%%%%%%%%%%%%%%%%%%%%%%%%%%
%\abbreviations{IR,NMR,UV}
%\keywords{American Chemical Society, \LaTeX}

%%%%%%%%%%%%%%%%%%%%%%%%%%%%%%%%%%%%%%%%%%%%%%%%%%%%%%%%%%%%%%%%%%%%%
%% The manuscript does not need to include \maketitle, which is
%% executed automatically.
%%%%%%%%%%%%%%%%%%%%%%%%%%%%%%%%%%%%%%%%%%%%%%%%%%%%%%%%%%%%%%%%%%%%%
%\begin{document}


%\begin{tocentry}
%\centering
%\includegraphics[width=6cm]{toc.eps}
%\label{For Table of Contents use only}
%\end{tocentry}

\begin{abstract}

Protein RNA-binding domains selectively interact with specific RNA
sites, a key interaction that determines the emergent cooperative
behaviors in RNA-protein mixtures. Through molecular dynamics
simulations, we investigate the impact of the specific binding
interactions on the phase transitions of an examplary RNA-protein
system and compare it with predictions of the Semenov-Rubinstein
theory of associative polymers.  Our findings reveal a sol-gel
(percolation) transition without phase separation, characterized by
double reentrant behavior as the RNA or protein concentration
increases. We highlight the crucial role of bridge formations in
driving these transitions, particularly when binding sites are
saturated. The theory quantitatively predicts the binding numbers at
equilibrium in the semidilute regime, but it significantly
overestimates the size of the concentration range where percolation is
observed. This can \fsrev{partly} be traced back to the fact that the
\fsrev{mean-field assumption in the} theory \fsrev{is not valid in the
dilute regime, and that the theory} neglects the existence of cycles
in the connectivity graph of the percolating cluster at the sol-gel
transition. Our study enriches the understanding of RNA-protein phase
behaviors, providing valuable insights for \fsrev{the interpretation
of} experimental observations.

\end{abstract}

\maketitle

\section*{I. Introduction}
% \section{Introduction}

Phase transitions play a crucial role in maintaining the structural
integrity and functional diversity of biological
systems\cite{hyman2014liquid}. Currently, liquid-liquid phase
separation (LLPS) is widely studied in cellular biophysics, especially
with respect to its potential importance for the formation of
membraneless organelles.  The complex interplay of protein and RNA
interactions leads to a diverse spectrum of phase behaviors governed
by factors such as temperature, pH, salt conditions, and the molecular
architecture\cite{banani2017biomolecular}.  Most research on LLPS has
focused on protein phase separation and its role in organelle
formation\cite{brangwynne2009germline}.  However, many cellular
processes such as RNA splicing, transport, and stability, depend on
the combined action of RNA and proteins\cite{bolognani2008rna,
wickramasinghe2016rna, shi2017mechanistic,
rahmanto2023k6,poblete2021rna}.  

While extensive research has been devoted to understanding phase
separation in RNA-protein
systems\cite{lin2015formation,maharana2018rna}, highlighting RNA's
ability to enhance phase separation alongside protein-protein
interactions\cite{molliex2015phase}, gelation, which is another
important cooperative phenomenon,  has received much less attention.
The sol-gel (percolation) transition describes the transformation of a
liquid-like sol into a gel-like state, characterized by the formation
of interconnected networks that span the entire system.  Already in
systems containing only proteins, gelation without phase separation
has been observed, pointing at a complex interplay of molecular
interactions\cite{harmon2017intrinsically,ranganathan2020dynamic}.
Introducing RNA into these systems adds another layer of complexity
due to the asymmetry between RNA and protein components, resulting in
an even  wider spectrum of phase
behaviors\cite{jankowsky2015specificity,garcia2019rna}. 
This highlights the necessity for more research into how
specific RNA-protein interactions lead to the formation of
structured networks.

The structural diversity of RNA and proteins --  such as RNA's linear
sequences, complex tertiary structures, and proteins' functional
domains like intrinsically disordered regions (IDRs) and RNA-binding
domains (RBDs) -- plays a crucial role in shaping phase behavior. The
solubility of proteins, often influenced by IDRs, and the binding
specificity provided by RBDs, are important factors for the formation
of condensates and gels \cite{harmon2017intrinsically,
martin2020intrinsically}.  RBDs, in particular, determine how proteins
interact with various RNA sequences, leading to the formation of
different phase-separated structures\cite{hentze2018brave,
gebauer2021rna}.  Different RNA species have been shown to induce
different condensate properties in RNA-protein mixtures. For example,
the interactions between Poly(A), Poly(U), and Poly(C) and proteins
typically give rise to the formation of liquid-like droplets, whereas
Poly(G)-protein mixtures form gel-like structures
\cite{roden_rna_2021,boeynaems_spontaneous_2019}.  Understanding the
molecular origin of these different behaviors is thus essential for
unraveling the complex cooperative phenomena observed in RNA-protein
mixtures.

The interactions between the binding sites in RNA and proteins are
traditionally classified as either specific or nonspecific \cite{jankowsky2015specificity, lin2015formation,
zhang2015rna, molliex2015phase,kang2019unified}.  Specific
binding involves the interaction of RBDs in proteins with
special sequences or motifs on RNA, whereas  nonspecific
binding enables proteins to interact with RNA sites
lacking identifiable sequences or structural motifs. These
interactions encompass a range of forces, including charge–charge,
dipole–dipole, $\pi$–$\pi$, and cation–$\pi$
interactions\cite{brangwynne2015polymer, protter2018intrinsically,
qamar2018fus,alberti2019considerations}.  Electrostatics is often
one of the key factors driving or controlling LLPS, but 
some studies have also identified phase transitions within
complex biological systems that mostly
rely on non-electrostatic interactions
\cite{harmon2017intrinsically, choi_lassi_2019,
ranganathan2020dynamic, choi2020generalized}.

To model such systems in molecular dynamics simulations, two
prevailing approaches have emerged.  Sticker-spacer models utilize
''stickers'' to mimic the multivalent binding domains on RNA or
protein chains, which are connected by neutral domains called
spacers\cite{semenov1998thermoreversible, harmon2017intrinsically,
choi_lassi_2019, borcherds2021intrinsically}.  Alternatively,
patchy-particle models treat each multivalent chain as a sphere with
different kinds of anisotropic attractive patches on its
surface\cite{espinosa_liquid_2020,joseph_thermodynamics_2021}.  These
two minimal models have proven effective in explaining the phase
behavior of biological systems. In the patchy-particle models, one can
control the number of patches on the surface of the particles to mimic
the multivalent character.  For example, Espinosa and coworkers have
used a patchy particle model to demonstrate that higher valencies
result in LLPS, and condensate formation is suppressed if the valency
is 2 \cite{espinosa_liquid_2020}.  In the sticker-spacer models,
multivalency is characterized by the number of stickers on a chain.
Harmon and colleagues demonstrated the critical role of spacer
solubility in influencing phase
behavior\cite{harmon2017intrinsically}. They found that in associative
polymers with the same structure, the explicit inclusion of
spacers—occupying significant volume—prevents phase separation.

In the present study, we explore the cooperative effects that specific
binding interactions induce in RNA-protein solutions.  We focus on
specific binding effects characterized by exclusive interactions
between a sticker from RNA and a sticker from a protein, thereby
preventing engagement with additional stickers. To this end, we
consider an idealized minimal spring-bead model with beads that have
no non-specific interactions apart from their excluded volume
interactions.  The emerging cooperative behavior in this model is
studied using two different approaches: First, large-scale off-lattice
molecular dynamics simulations, and second, an extension of the
Semenov-Rubinstein theory of percolation and phase separation
\cite{semenov1998thermoreversible} to asymmetric heteroassociative
mixtures \cite{danielsen2023phase}.  We find that specific binding may
lead to percolation without phase separation, resulting in the
formation of connected clusters spanning across the whole system.
Moreover, our study reveals the possibility of double-reentrant
sol-gel-sol transitions upon varying the concentrations of RNA and
proteins, in agreement with theoretical expectations
\cite{danielsen2023phase}.  To gain insights into the percolation
transitions, we quantify the frequency of different binding structures
such as bridges, which are central for network formation, loops, and
isolated bonds.  A comparison between theoretical predictions and
simulation results shows that the theory captures the phase behavior
and fairly accurately predicts binding numbers, particularly at high
concentrations \fsrev{in the semidilute regime}. Nevertheless, the
theory falls short in quantitatively predicting percolation threshold
values and overestimates the size of the parameter region where the system is gelated. 

%This discrepancy is
%attributed to local correlations between the reversible bonds
%\xcrev{and the excluded volume effect in the low concentration}, which
%are neglected in the mean-field theory.
%FS: I would just remove this sentence. 

\section*{II. Model and methods}

\subsection*{II.A Simulation}
\label{sec:simulation}

\begin{figure}[ht]
\includegraphics[width=8cm]{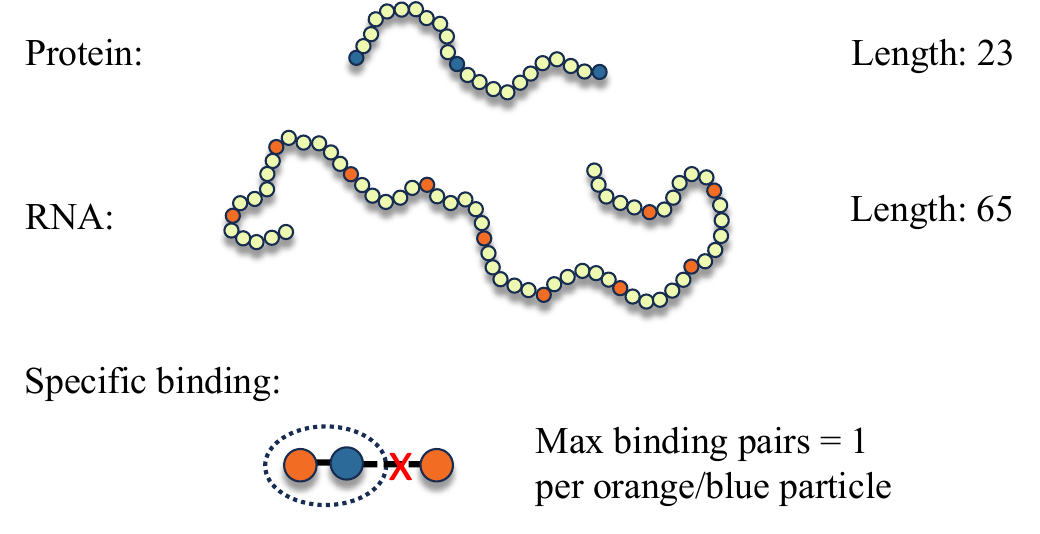}

\caption{Cartoon showing the model for RNA and protein in the
present study. Blue, orange, and yellow beads are stickers in protein
and RNA chains and spacers, respectively. To account for the
specificity of binding interactions, each protein sticker
is designed such that it can pair exclusively with a single RNA sticker.
}

\label{fgr:cartoon}
\end{figure}

We model RNA molecules and proteins in good solvent using a
sticker-spacer model.  Specifically, we consider proteins with
distinct functional motifs capable of associating with certain RNA
sequences, mirroring the finite number of RNA-binding domains (RBDs)
found in natural proteins. For instance, hnRNPH, an  RNA-binding
protein within the heterogeneous nuclear ribonucleoproteins family,
possesses three RNA-binding motifs\cite{krecic1999hnrnp}, which play a
crucial role in modulating the structure of RNA G quadruplexes, as
highlighted in recent
studies\cite{decorsiere2011essential,herviou2020hnrnp}.  Inspired by
this, the ''proteins'' in our simulation also contain 3 binding
domains, and we will refer to them as trimers in the following.  RNA
molecules will be modeled as long \fsrev{spring-bead} chains with
several binding sites.  Binding sites will be modeled as stickers,
which are connected by purely repulsive spacer chains. A schematic
cartoon of these chain structures is shown in Fig.
\ref{fgr:cartoon}.  Blue and orange beads correspond to stickers in
protein and RNA chains respectively. Yellow beads correspond to
spacers, linking stickers along each chain.  RNA and protein stickers
can associate with each other, mimicking specific binding, but
stickers of the same type repel each other.  Moreover, each sticker
can only associate with at most one other sticker.

Putting this more formally, we consider a system of $n_A$ polymer
chains $A$ (RNA) $n_B$ polymer chains $B$ (proteins/trimers) in a
volume $V$ at temperature $T$.  Polymers of type $i$ ($i=A,B$) contain
$N_i$ monomers (\fsrev{beads}), which includes $f_i$ stickers that are
evenly distributed along the chain. The mean monomer density of chains
$i$ is thus $c_i = n_i N_i/V$, and the mean density of $i$-stickers is
$c_{si} = n_i f_i/V$.  For simplicity, all monomers are taken to have
the same mass $m$ and diameter $\sigma$. Adjacent monomers in a chain
are connected by a harmonic bond potential: 
\begin{equation}
\label{eq:bond}
\beta U_{\text{bond}}=\frac{1}{2}k_b(r-r_0)^2
\end{equation}
Here $\beta=1/k_BT$  is the Boltzmann factor,
$k_b$ is the spring constant, $r_0=\sqrt[6]{2} \ \sigma$ is the
equilibrium bond length and $\sigma$ is the diameter of the
monomers as mentioned above, which serves as the length unit in our
paper.  Non-bonded pairs of monomers with the exception of
pairs of A- and B-stickers interact via the purely repulsive
Weeks–Chandler–Andersen (WCA) potential\cite{weeks1971role}:
\begin{equation}
\label{eq:wca}
\beta U_{\text{WCA}}=\left\{
\begin{array}{rcl}
&4\epsilon\left[\left(\frac{\sigma}{r}\right)^{12}-\left(\frac{\sigma}{r}\right)^6+\frac{1}{4}\right] & {r<\sqrt[6]{2}\sigma} \\
&0& {r\geq \sqrt[6]{2}\sigma}
\end{array}\right.,
\end{equation}
where $\epsilon$ is the interaction strength of the WCA potential. To
emulate the specific binding between RNA and proteins, we incorporate
a targeted interaction between $A$ and $B$ stickers. It ensures that a
single $A$-sticker is capable of binding with just one $B$-sticker and
vice versa, while any additional stickers are subject to exclusion
through the WCA potential, as enforced by the already bound partner
sticker.  The specific binding potential
is\cite{zhang2021decoding,grandpre2023impact}:
\begin{equation}
\label{eq:sp_inter}
\beta U_{\text{binding}}=\left\{
\begin{array}{rcl}
&-\eb\left[\cos\left(\frac{2\pi r}{\sigma}\right)+1\right] & {r<0.5\sigma} \\
&0& {r\geq 0.5\sigma}
\end{array}\right.
\end{equation}
The strength of specific binding between stickers in RNA and trimers
is controlled by $\eb$. In real RNA-protein systems, the RNA binding
domains in proteins can link to RNA dynamically and reversibly through
a weak chemical binding interaction, hence $\eb$ can not
be very large.  Importantly, the specific binding potential's cut-off
distance ($0.5\sigma$) combined with the purely repulsive interactions
between homotypic monomers ensures the specificity of binding (see
Fig. \ref{fgr:cartoon}).  The implementation of specific binding via
a simple pairwise and isotropic attractive potential has distinct
advantages from a computational point of view, because it can be
implemented in a straightforward manner using standard optimized
simulation packages.  An alternative and arguably more natural way to
mimic reversible directed binding between protein and RNA domains
would be to implement reversible reactions, with Monte Carlo moves
that satisfy detailed balance. Our model can capture this
reversibility by controlling the binding process at the particle level
through the competition between different interactions within the
system.

\begin{figure}[ht]
\includegraphics[width=8cm]{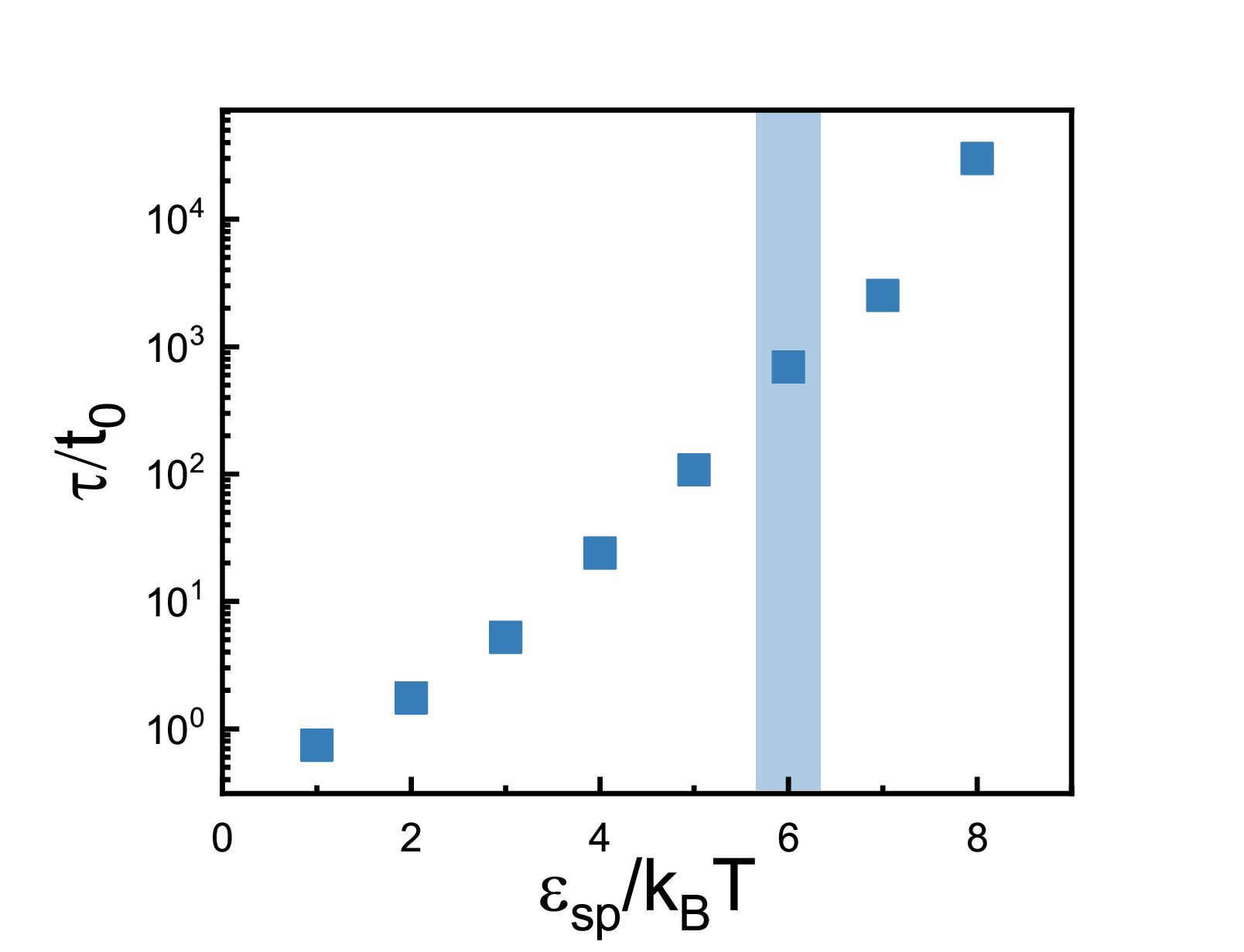}
\caption{The lifetime of specific binding for different binding
strengths $\eb$, as obtained from simulations of
one RNA and one trimer molecules, see text, in a small cubic
box of side length $20 \sigma$.
}
\label{fgr:tau}
\end{figure}

We perform Langevin dynamics simulations with implicit solvent at a
fixed temperature ($k_B T$ is the unit of energy), using the
simulation package HOOMD-blue(version
2.9.3)\cite{anderson2008general}. The snapshots of our systems are
visualized with  OVITO\cite{ovito}.  The chain lengths of RNA and
trimers are $N_A=65$ and $N_B=23$, respectively, and the number of
stickers  in each chain is $f_A=10$ and $f_B=3$. \fsrev{They are separated
by spacers of length $l_A=5$ and $l_B=10$}. The spring
constant is set to $k_b=30k_B T/\sigma^2$\fsrev{, resulting in 
a statistical segment length of $a\approx1.2 \sigma$}. We simulate the RNA-trimer
system in a rectangular box ($100\sigma\times L_y\times 80\sigma$), in
which the box size can be changed along the $y$-axis direction. The
total duration of simulation runs is $10^6 t_0$ to make sure the
system reaches the equilibrium state, with $t_0= \sqrt{m \sigma^2/k_B
T}$ being the time unit of the simulation. The strength of the WCA
potential $\epsilon=1 \: k_B T$. To specify $\eb$, we ran separate
simulations of a system containing only one RNA and one trimer, and
determined the average life time of specific binding as a function of
$\eb$. The results are shown in Fig. \ref{fgr:tau}.  At $\eb=6 k_B
T$, we obtain $\tau\sim 10^3 t_0$, implying that the specific bonds
can open and close frequently during one simulation run. Assuming that
our simulation time unit corresponds to a real time span of the order
of a picosecond, this lifetime scale aligns well with reversible
binding events observed in real biological
systems\cite{galvanetto2023extreme}.  Therefore, in the following
discussion, we keep $\eb=6k_B T$. Error bars on averaged quantities
were estimated by computing the standard error of the mean.

\subsection*{II.B Theory}
\label{sec:theory}
 
In Section III,
%\ref{sec:results}
we will compare the simulation results with the predictions of the
Semenov-Rubinstein theory for reversible gelation in solutions of
associative polymers \cite{semenov1998thermoreversible,
prusty2018thermodynamics, choi2020generalized, michels2021role,
danielsen2023phase}.  Therefore, we now briefly recapitulate the
central ideas of this mean-field approach and the main equations for
the case of our particular system.  We consider a two-component system
of $n_A$ and $n_B$ chains with length $N_A$ and $N_B$, each containing
$f_A$ or $f_B$ stickers, in good solvent. $A$ and $B$ stickers can
form reversible bonds, thereby gaining an energy ($- \eb$) per bond.
\fsrev{The mean-field approximation consists in neglecting the
correlations between sticker positions and chain conformations, such
that the stickers behave like an ideal gas. This implies that
the solution is semidilute in the sense that the spacers between
stickers overlap, and that the influence of excluded volume interactions 
on sticker-sticker contacts can be 
neglected\cite{semenov1998thermoreversible,danielsen2023phase}.
It also implies that the chains are long and that the number of
stickers per chain is high.
} The free energy per volume of the whole system is \fsrev{then} composed of
three parts: 
\begin{equation}
\label{eq:f_total}
\beta f=\beta f_\mr{entropy}+\beta f_\mr{sticker}+\beta
f_\mr{int}.
\end{equation}
The first term is the pure translational entropy of
polymer chains as a function of the monomer concentration
$c_i=n_iN_i/V$:
\begin{equation}
\label{eq:f_entropy}
\begin{aligned}
\beta f_\mr{entropy}&=\frac{c_A}{N_A}\ln\left(\frac{c_A}{N_A e}\right)+\frac{c_B}{N_B}\ln\left(\frac{c_B}{N_B e}\right).
\end{aligned}
\end{equation}
The second term, $\beta f_\mr{sticker}$, is the free energy
associated with the formation of reversible bonds between $A$- and
$B$-stickers.
For a given number $N_p$ of bonds, it is calculated as follows:
\begin{equation}
\label{eq:f_sticker_1}
\begin{aligned}
\beta f_\mr{sticker}&
  =-\frac{1}{V}\ln\left[P_\mr{comb}
  \left(\frac{v_b}{V}\right)^{N_p}e^{\beta \eb N_P} \right],
\end{aligned}
\end{equation}
where $v_b$ is the bond volume for specific binding and
$P_\mr{comb}$ is the number of different ways how to distribute
the $N_p$ ($AB$) bonds onto $f_A n_A$ stickers of type $A$ and $f_B
n_B$ of type $B$:
\begin{equation}
\label{eq:P_comb}
\begin{aligned}
P_\mr{comb}=\tbinom{n_Af_A}{N_p}\tbinom{n_Bf_B}{N_p}(N_P)!.
\end{aligned}
\end{equation}
Using Stirling's approximation, 
Eq.\ (\ref{eq:f_sticker_1}) can be rewritten as
\begin{equation}
\label{eq:f_sticker_2}
\begin{aligned}
\beta f_\mr{sticker}& 
   =\frac{N_p}{V}+\frac{n_Af_A}{V}\ln(1-p_A)
     +\frac{n_Bf_B}{V}\ln(1-p_B)\\
    &-\frac{N_p}{V}\ln\left[\frac{(n_Af_A-N_p) (n_Bf_B-N_p)}{N_p}
       \frac{K^{-1}}{V}\right],
\end{aligned}
\end{equation}
in which $K=v_b^{-1} e^{-\beta \eb}$ is the dissociation constant.
Here $p_i=N_p/n_i f_i$ ($i=A,B$) is the fraction of bound stickers for
each component $i$. 

The last term in Eq.\ (\ref{eq:f_total}) accounts for the
excluded volume interactions between all monomers except $A$-
and $B$-stickers,
\begin{equation}
\label{eq:f_int}
\begin{aligned}
f_\mr{int}=
\frac{1}{2} v \left[ (c_A+c_B)^2 - 2 \csa \: \csb \right],
\end{aligned}
\end{equation}
which depends on the excluded volume parameter $v$ and the $i$-sticker
concentrations $c_{si} = n_i f_i/V = c_i f_i/N_i$ ($i=A,B$). 
In the following, we will characterize the system 
in terms of the sticker concentrations $c_{si}$ instead
of monomer concentrations $c_i$ for convenience, and
also introduce the total sticker concentration 
$c_s = \csa + \csb$.  Minimizing the free energy with respect 
to $N_p$ for given $c_{si}$ we obtain
\begin{eqnarray}
\label{eq:N_p_eq}
N_p & = & 
\frac{V}{2}(K+ c_s - \sqrt{\Delta}),
\\ \nonumber
\text{with} &&
\Delta = (K+ c_s)^2 - 4 \csa \csb,
\end{eqnarray}
which allows to calculate
\begin{equation}
\label{eq:p_i}
p_i = N_p/n_i f_i = N_p/(c_{si}V),
\end{equation}
and to derive the following expression for the free energy per
area as a function $\csa,\csb$:
\begin{equation}
\label{eq:f}
\begin{aligned}
\beta f&=
\frac{\csa}{f_A}\ln\left(\frac{\csa}{f_A e}\right)
+\frac{\csb}{f_B}\ln\left(\frac{\csb}{f_B e}\right)\\
&+\csa  \left[\frac{p_A}{2}+\ln\left(1-p_A\right)\right]\\
&+\csb \left[\frac{p_B}{2}+\ln\left(1-p_B\right)\right]\\
&+\frac{1}{2}v \left[\left(\frac{\csa N_A}{f_A} + \frac{\csb N_B}{f_B}\right)^2
   - 2\csa \csb\right].
\end{aligned}
\end{equation} 
Based on this expression, the stability and phase behavior of the
system can be analyzed by standard methods. \fsrev{According to
Danielsen {\em et.\ al.}\cite{danielsen2023phase}, the underlying mean-field
assumption is expected to become problematic if $vl > (a\sqrt{l})^3$,
(excluded volume effects become important at the spacer level), and/or
if $c_{si} < (a \sqrt{l})^{-3}$ for $i=A$ or $B$ (spacers do not
overlap), where $l$ is the spacer length and $a$ the statistical
segment length.}

Owing to the formation of specific bonds between polymers, the
multivalent polymer chains can connect to a system-spanning network
structure once the concentration reaches a certain threshold. This
threshold for the sol-gel transition can be estimated using analytical
approaches derived from the Flory-Stockmayer theory, which provides a
basic understanding of polymer network formation and gelation
processes\cite{flory1953principles,stockmayer1943theory,semenov1998thermoreversible,choi2020generalized,harmon2017intrinsically}.
Our system contains two different polymer components in solution. For
each component $i$, the number of binding sites along one chain is
$f_i$ and the fraction of binding sites that are occupied is $p_i$. 

We define clusters as sets of chains that are directly or
indirectly connected to each other by specific bonds, and consider the
thermodynamic limit $V \to \infty$. A gel is then characterized
by the existence of an infinite cluster.  A given chain of type $i$
that is connected to a given cluster by one bond has, on average,
$p_i(f_i -1)$ additional bonds. Now let us consider a hypothetical
algorithm that identifies clusters in a given configuration of $(AB)$
bonds by reconstructing them in a stepwise fashion, starting from a
single initial chain.  In our system, $A$-stickers can only bind to
$B$-stickers and vice versa.  Thus an $A$ chain that is connected to
a partially reconstructed cluster via one $B$-chain has on average
$p_A (f_A-1)$ additional connections to $B$-chains, and these
$B$-chains have on average $p_B (f_B-1)$ additional connections to
$A$-chains. Some of these connections will cycle back to the original
partially reconstructed cluster. Thus, the $A$-chain in question has
on average less than $m = p_A (f_A-1) p_B (f_B-1)$ $B$-mediated
connections to ''new'' $A$-chains, i.e., chains that are not yet part 
of the original reconstructed cluster. In order to reconstruct an infinite
cluster, $m$ must necessarily exceed one. In the Rubinstein-Semenov
theory \cite{semenov1998thermoreversible,danielsen2023phase}, the
possibility of cycles is neglected and it is assumed that every
additional connection truly extends the size of the reconstructed
cluster. In this case, we expect the percolation transition at
\begin{equation} 
\label{eq:P_c_theory} 
\begin{aligned}
m =  (f_A-1)p_A\times (f_B-1)p_B \stackrel{!}{=} m_p = 1 .
\end{aligned}
\end{equation}
\xcrev{We emphasize that this condition is independent of the way how
the $p_i$ has been obtained. It is derived based on the assumption
that graphs are tree-like, without accounting for specific molecular
characteristics of the particular system. The threshold predicted by
this equation only depends on the sticker number $f_i$ and binding
probability $p_i$. Therefore, the}
%The binding probability itself can originate from 
%any specific system and may encapsulate information
%unique to that system.} 
%FS: I don't understand this sentence, therefore I removed it.
criterion must be seen as a rigorous lower bound for the value of $m$
at the percolation threshold \fsrev{for given $p_i$}. The true threshold 
value will be higher. 

\section*{III Results and discussion}
\label{sec:results}

%\subsection{Phase diagram}
\subsection*{III.A Stability of the homogeneous system}
\label{sec:results_stability}

\begin{figure*}[ht]
\includegraphics[width=13cm]{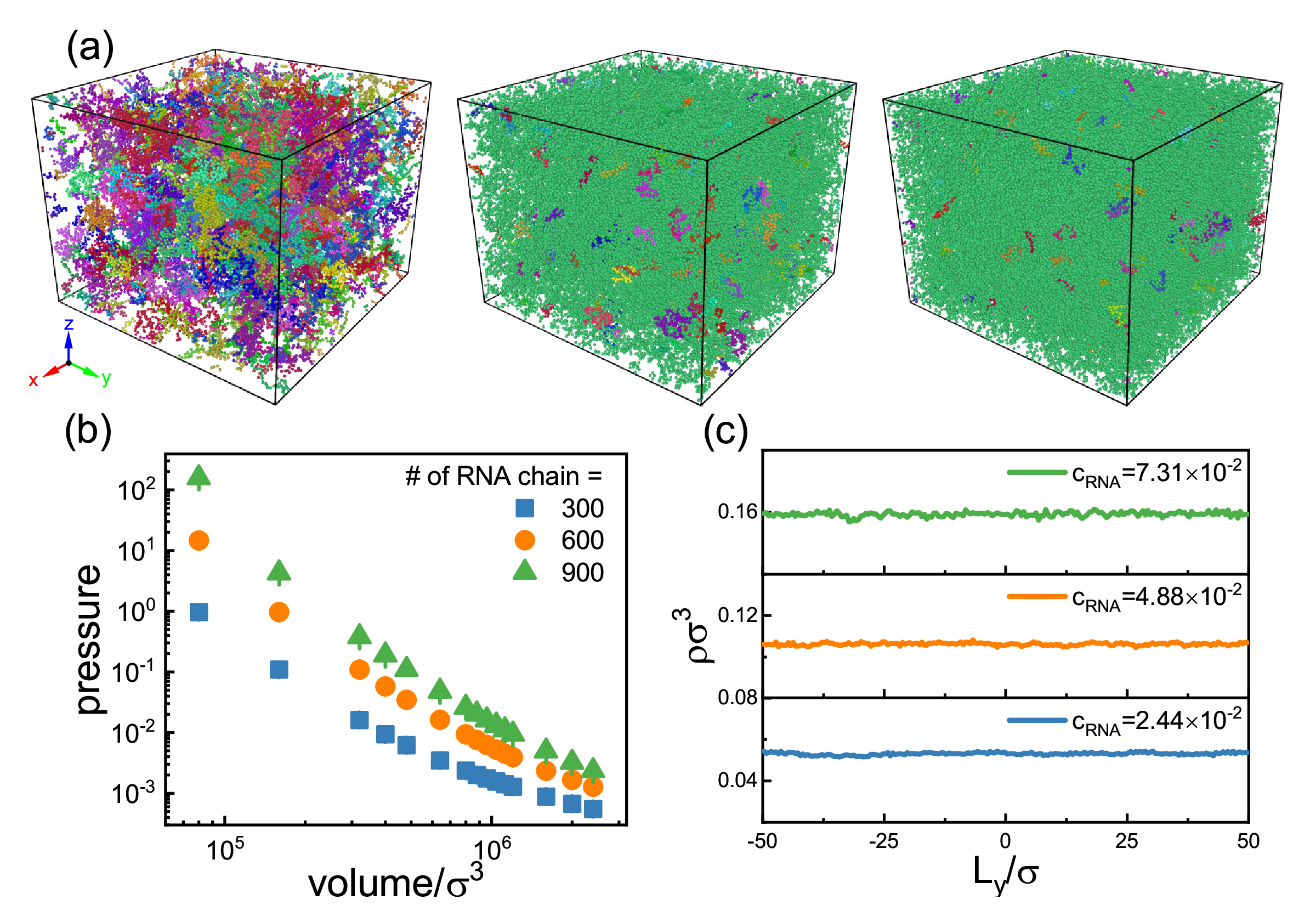}
\caption{(a) Snapshots of the system
for different \fsrev{RNA monomer} concentrations in the equilibrium state
(Left to right: $c_\mr{RNA}=2.44\times 10^{-2}\sigma^{-3}, 4.88\times
10^{-2}\sigma^{-3}, 7.31\times 10^{-2}\sigma^{-3}$).  Multi-colored
particles belong to different clusters. (b) Pressure-volume graph
for different numbers of RNA chains.
In each case, the ratio of RNA and trimer chains is fixed at
$n_\mr{RNA}/n_\mr{trimer}=3/10$, such that the total numbers of
stickers of RNA and trimers match each other.
%are the same, which means the ratio 
(c) Average density profile along $y$ direction for each
case in the volume $8\times 10^{5}\sigma^{3}$.} 
\label{fgr:pd}
\end{figure*}

In the exploration of the behavior of the RNA-trimer system, we begin
with investigating the stability of the homogeneous system.  This is
first \pb{done} analytically using the theory described above.  Within
this theory, a homogeneous mixture is (meta)stable if the Hesse matrix
of the free energy per volume $f$ (Eq. (\ref{eq:f})) as a function of
$\csa$ and $\csb$ , $H_{f;ij} = \partial^2 f/\partial c_{si} \partial
c_{sj}$, is positive definite. The diagonal elements of the Hesse
matrix are given by
\begin{equation}
\frac{\partial^2 \beta f}{\partial c_{si}^2}
= p_i/\sqrt{\Delta} + v (N_i/f_i)^2 + 1/(f_i c_{si}) > 0,
\end{equation}
(using the definitions of Eq.\ (\ref{eq:N_p_eq})) and always positive.
However, the determinant may become negative. In the extreme case of 
very strong binding $(K \to 0)$ and zero excluded volume interactions 
($v=0$), the determinant takes the value
\begin{eqnarray}
&&\det(H_f)\big|_{v=0, K\to 0}
= \frac{1}{\csa \csb}  \times
\nonumber \\ && \left[ \frac{1}{f_A f_B} 
 -  \frac{1 - \frac{1}{f_A} - \frac{1}{f_B}}{2}
\Big(\frac{c_s}{|\csa - \csb|} - 1\Big) \right].
\label{eq:det_v0K0}
\end{eqnarray}
This result shows that the homogeneous state necessarily becomes unstable 
for $f_A,f_B > 2$ in the region where the concentrations $\csa$ and 
$\csb$ of $A$- and $B$-stickers roughly match, $\csa \approx \csb$. 

Excluded volume interactions can restore the stability. For $v>0$ and
$\csa \approx \csb \approx c_s/2$, the determinant is
\begin{eqnarray}
\det(H_f)\big|_{K\to 0}
&\approx&  \frac{2}{|\csa - \csb|} \Bigg[
- \frac{1 - \frac{1}{f_A} - \frac{1}{f_B}}{c_s}
\nonumber \\ &&
+ \frac{1}{2}\: v \left( \left(\frac{N_A}{f_A}+ \frac{N_B}{f_B}\right)^2 - 2 \right)\Bigg],
\label{eq:det_vK0}
\end{eqnarray}
where we have omitted terms that do not scale like $1/|\csa - \csb|$.
This equation shows that the excluded volume interactions
can stabilize the homogeneous state at higher concentrations. The
effect increases with increasing spacer content in the system, i.e.,
large values of $N_i/f_i$. Nevertheless, the system will remain
unstable at low concentrations, i.e., $c_s \to 0$ as long as
$K \to 0$, i.e., at infinite binding strength. 

If the binding strength is finite, bond formation
is suppressed in the very dilute regime for entropy reasons, which
also restores stability. For $K>0$ and $v=0$, the determinant of the
Hesse matrix for small total sticker concentration
$c_s$ reads
\begin{eqnarray}
\det(H_f)\big|_{v=0}
&=& \frac{1}{\csa \csb} \Bigg[ \frac{1}{f_A f_B}
+ \frac{1 - \frac{1}{f_A} - \frac{1}{f_B}}{2} \times
\\ \nonumber &&
\left(
1 - \frac{K + c_s}{\sqrt{(K+c_s)^2-4\csa\csb}}
\right)
\Bigg]
\\ \nonumber & \approx &
 \frac{1}{f_A f_B \csa \csb} - \frac{1 - \frac{1}{f_A} - \frac{1}{f_B}}{K^2}
 + {\cal O}(c_s).
\label{eq:det_v0K}
\end{eqnarray}
For sufficiently large dissociation constant $K$, the instability
at small concentrations is thus removed.

In order to apply the theory to the simulation model, we must
determine the excluded volume parameter $v$. This parameter, also
known as the second virial coefficient, is calculated using the
Mayer-f function:
\begin{equation}
\label{eq:v_0}
v_\mr{ex}=2\pi\int_0^\infty \text{d} r  \: r^2\left(1-e^{-\beta w(r)}\right).
\end{equation}

Here, $\beta w(r)$ represents the nonbonded monomer-monomer
interaction potential, i.e., the WCA potential (Eq.\ (\ref{eq:wca})).
Substituting the WCA potential in Eq.\ (\ref{eq:v_0}), we obtain
$v=2.2 \sigma^3$.  Inserting this value and $\eb = 6 k_B T$, we find that the
determinant of the Hesse matrix is always positive in our system,
indicating that the homogeneous state is stable or at least metastable.
A numerical evaluation of $f$ (Eq.\ (\ref{eq:f})) suggests that the
homogeneous phase is truly stable. Therefore, the theory predicts that
our system should not exhibit phase separation.

We emphasize that this is not a general result, but a \fsrev{feature}
of the \fsrev{system} studied in the present work. The
Semenov-Rubinstein theory of associative polymer solutions does
predict the existence of a phase-separated
regime\cite{semenov1998thermoreversible, choi2020generalized,
harmon2017intrinsically}. Danielsen et al.\ have shown that the extent
of this regime is significantly reduced in mixtures of
heteroassociative polymers, compared to single-component systems of
associative polymers \cite{danielsen2023phase}. The considerations
above highlight the role of excluded volume interactions between the
spacer monomers in suppressing phase separation at high polymer
concentrations, in agreement with arguments by Harmon et al.\
\cite{harmon2017intrinsically}, and the role of the finite binding
strength in suppressing phase separation at low polymer
concentrations. 

To verify the theoretical prediction in simulations, we focus on the
case $\csa = \csb$ where instabilities are most likely to occur in the
homogeneous mixture. Therefore, we set the ratio of $A$-chains (RNA)
to $B$-chains (trimers) to $n_A/n_B= 3/10$, such that $f_A n_A = f_B
n_B$. Fig. \ref{fgr:pd}(a) shows snapshots of our system in the
equilibrium state for three different polymer concentrations.
Polymers belonging to different clusters are colored differently.
With increasing concentration, RNA and trimer chains bind together to
form larger clusters. At low concentrations, (\fsrev{RNA bead
concentration} $c_\mr{RNA}=2.44\times
10^{-2}\sigma^{-3}$), small clusters are uniformly dispersed
throughout the system (see Fig. \ref{fgr:pd}(a), left panel),
whereas at higher concentrations, a single connected cluster spans the
whole system (Fig. \ref{fgr:pd}(a), middle and right panel).
Nevertheless, no phase separation occurs. To demonstrate this, we
utilize two kinds of verification methods.  According to
thermodynamics, phase separation is associated with a plateau in a
graph of pressure versus the volume at fixed molecule number.  In
small systems, this plateau could be replaced by nonmonotonic behavior
(a van-der-Waals loop).  Fig. \ref{fgr:pd}(b) shows the
pressure-volume relationships for three different values of polymer
numbers.  The pressure decreases monotonously with increasing volume,
with no signature of unusual behavior at any concentration.
Alternatively, inspecting the density profile offers another method to
investigate the phase behavior directly.  This method is often
utilized in recent research
\cite{benayad2020simulation,rekhi2023role}.  The density profiles
along the $y$-direction are shown in Fig. \ref{fgr:pd}(c). They are
nearly uniform, with no sign of large density variations as would be
characteristic for phase separation.  
%\xcrev{In our simulation, the excluded volume parameter $v=2.2$,
%means that our system is in a good solvent. While the
%Semenov-Rubinstein
%theory\cite{semenov1998thermoreversible,danielsen2023phase} is not
%properly valid in good solvent condition, from the theoretical
%discussion in Section III.A, the phase separation is also not
%expected in our system that supports our simulation results.
%Furthermore, in Eq.\ (\ref{eq:det_vK0}), the sign of the determinant
%depends on not only the excluded volume parameter but also the
%sticker number $f_i$ and the spacer ratio $N_i/f_i$. Therefore, the
%phase separation may happen in some specific chain structure. In our
%ongoing work, we indeed observe the phase separation in simulation.} 
In summary, the findings from the two methods strongly support the
theoretical result that our \xcrev{current} system does not experience
phase separation.  \fsrev{However, this is not a generic property of
our model. If we reduce the spacer length, i.e., reduce $N_i/f_i$, the
system exhibits a phase-separated region both according to theory and
simulations. This case will be discussed elsewhere.}

\subsection*{III.B Sol-gel transition}
\label{sec:results_percolation}

\begin{figure}[ht]
\includegraphics[width=8cm]{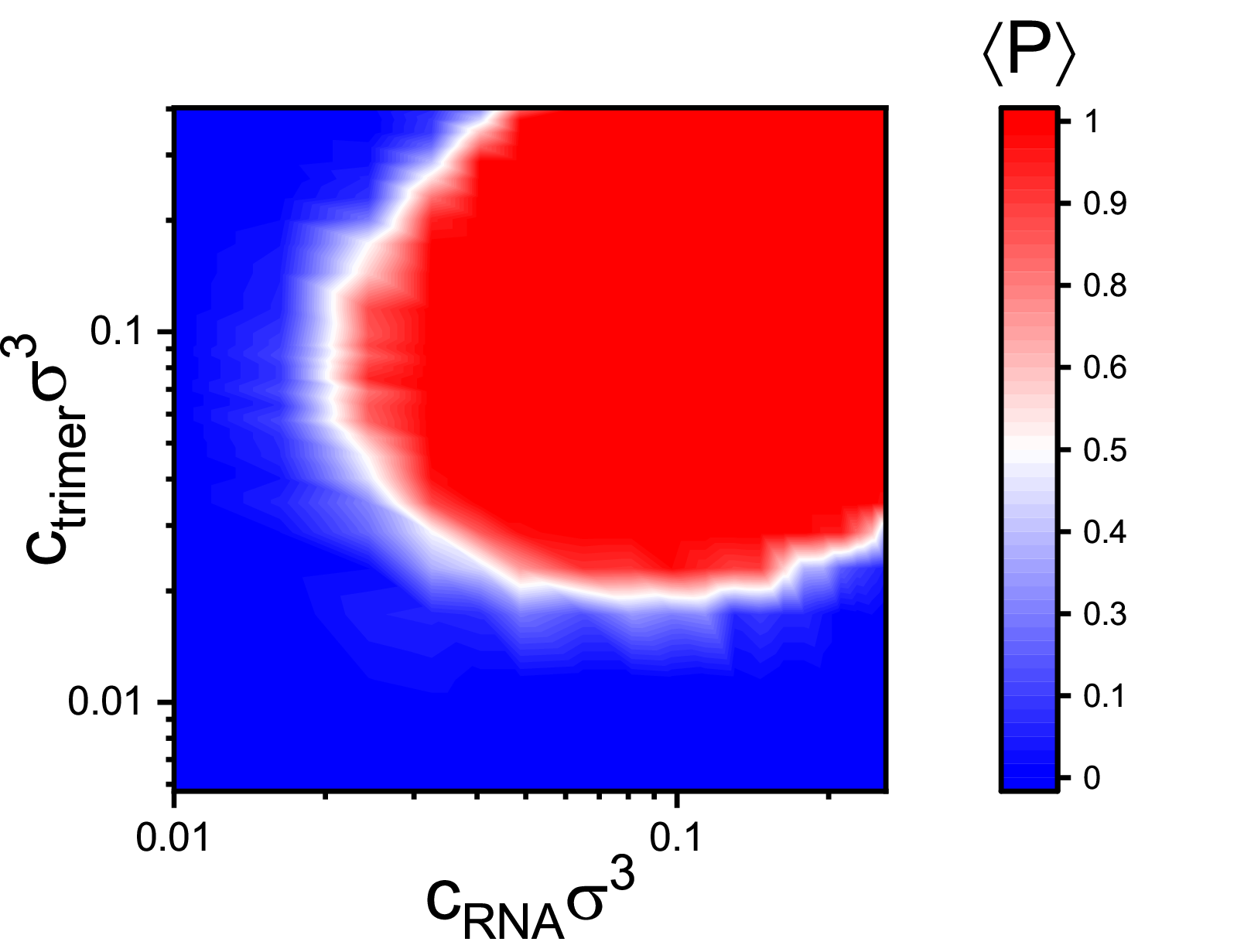}
\caption{Order parameter $\langle P \rangle$ with different
concentrations of RNA and trimers. In the red region, a percolation
transition occurs, with the white line indicating the percolation
threshold.}
\label{fgr:P_c}
\end{figure}

As discussed in Section II.B and in the
literature\cite{semenov1998thermoreversible, zeng2023developments},
systems of associative polymers may exhibit a percolation transition.
We recall that the term percolation denotes a geometrical transition
in an infinite system of partially connected units (polymers) between
a state where all connected clusters are finite to a state where a
large system-spanning cluster exists. Thus, the notion of
''percolation'' depends on the definition of connectivity in a
cluster. In many computational studies and cluster algorithms,
molecules are taken \pb{to} belong to the same cluster if they satisfy certain
proximity criteria, e.g., the minimum distance between monomers in the
two chains is below a given threshold value. Here, following the
spirit of the Semenov-Rubinstein theory, we take two chains to be
connected if a specific bond has formed between them. Due to the high
energy of specific interactions, the existence of an infinite cluster
then also implies a change in rheological properties and the emergence
of rubber-like behavior on sufficiently short time scales, as is
characteristic for a sol-gel transition.

Percolation is a sharp and well-defined transition in the
thermodynamic limit, where the spanning cluster is infinitely large.
In finite homogeneous systems, the transition smoothens, but it can
still be characterized in terms of spanning clusters. Since we use
periodic boundary conditions in our simulations, we can define a
spanning cluster through the requirement that a closed path within the
cluster can be traced from any particle to its periodic image. Based
on this definition, we define an order parameter for the percolation
transition $P$ {\em via}
\begin{equation}
\label{eq:P_order}
P=\left\{
\begin{array}{rcl}
&1 &: {\text{\fs{configuration} contains $\geq 1$ spanning clusters}} \\
&0&: {\text{otherwise}}
\end{array}\right.
\end{equation}
\xcrev{and we calculate the time-averaged value of this quantity,
$\langle P \rangle$, in the equilibrated systems. Since we consider
reversible binding, this corresponds to a statistical average with
Boltzmann weight.} In infinite systems, the percolation transition is
characterized by a jump from $\langle P \rangle = 0$ to $\langle P
\rangle = 1$. In finite systems, the jump is replaced by a continuous,
but quite sharp transition. \fsrev{This is because the probability to
sample a configuration with $P=0$, i.e., a configuration without
spanning clusters, is never strictly zero in finite systems, and the
same holds for $P=1$ if the number of chains in the system is high
enough that they can form a spanning cluster.}
%\xcrev{In our simulation, due to the reversible binding, no spanning
%clusters are present in the whole system sometimes, resulting in
%transient $P=0$, while spanning clusters exist, leading to $P=1$. }
Fig. \ref{fgr:P_c} shows the behavior of $\langle P \rangle$ in our
system for different concentrations of RNA and trimers and thus traces
out the phase diagram of the percolation transition.

The figure demonstrates one striking feature of our system: For fixed
RNA or trimer concentration, increasing the concentration of the other
component leads to reentrant behavior.  For example, at fixed trimer
\fsrev{bead} concentration $c_{\text{trimer}} \sim 0.03
\sigma^{-\pb{3}}$, percolation sets in once the RNA \fsrev{bead}
concentration reaches $c_{\text{RNA}} \sim 0.03 \sigma^{-\pb{3}}$, but
is suppressed again at $c_{\text{RNA}} > 0.25 \sigma^{-\pb{3}}$.
Likewise, at fixed RNA \fsrev{bead} concentration $c_{\text{RNA}}
\sim 0.03 \sigma^{-3}$, percolation is only observed in a window of
trimer \fsrev{bead} concentrations, $0.03 \sigma^{-3} <
c_{\text{trimer}} < 0.23 \sigma^{-3}$.  Summarizing the results shown
in Fig. \ref{fgr:pd} and \ref{fgr:P_c}, we conclude that specific
binding in our RNA-trimer system induces a percolation transition
without phase separation if the concentrations of the components reach
a certain threshold, but are not too much in excess of each other.

\begin{figure*}[ht]
\includegraphics[width=13cm]{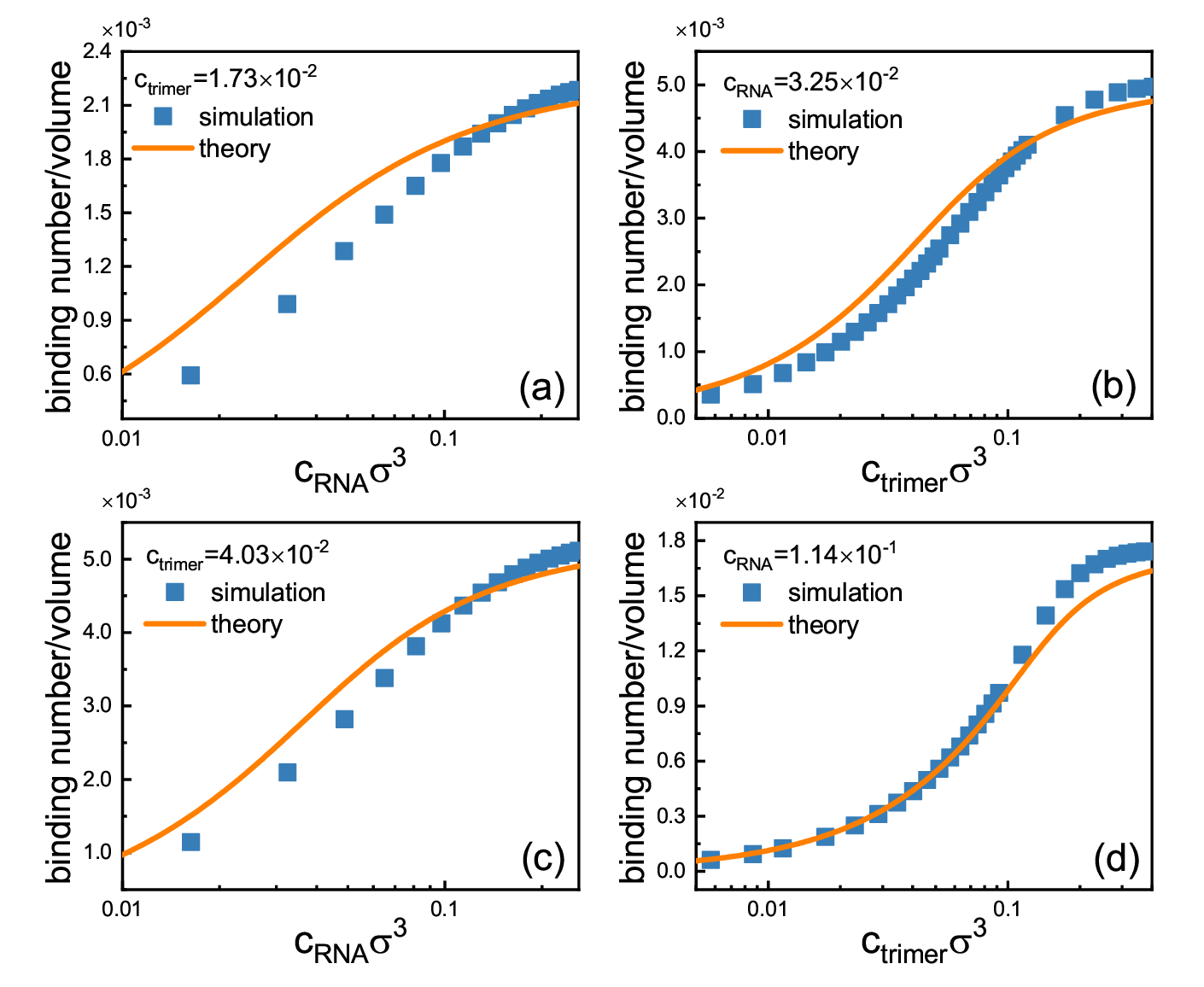}
\caption{\xcrev{Density} of specific bonds in the simulations (a,c) as
a function of RNA concentration for fixed trimer concentrations as
indicated and (b,d) as a function of trimer concentration for fixed
RNA concentrations as indicated.  Blue square symbols show the results
of the simulation. Orange solid lines are the results based on Eq.\
(\ref{eq:N_p_eq}) using the same parameters as in our simulation.}
\label{fgr:theory}
\end{figure*}

We will now set out to compare the simulation results with the theory
of Section II.B.
%\ref{sec:theory}. 
We start with considering the average
fraction of bound RNA and trimer stickers, $p_A$ and $p_B$, which are
calculated according to Eq.\ (\ref{eq:p_i}). Fig. \ref{fgr:theory}
shows the corresponding \xcrev{density of bound sticker pairs}, i.e., the
total number of specific bonds \xcrev{per volume, $N_p/V$}, for
different RNA and trimer concentrations, as obtained from simulations
and predicted by theory.  The theory agrees remarkably well with the
simulations \xcrev{especially at high concentrations, see Fig.
\ref{fgr:theory} (d)}, even though it has no adjustable parameters.
The largest discrepancies are observed at small trimer and RNA
concentrations, where the theory overestimates the binding numbers.
% FS: I removed the reference to 5b. 
%The largest discrepancies are observed at small trimer and RNA
%concentrations, where the theory overestimates the binding
%\xcrev{density}. 

\fsrev{As we have discussed earlier, central mean-field assumptions
in the Semenov-Rubinstein theory are that excluded volume interactions
are small on the spacer scale that spacers overlap. This results in
the conditions $v\cdot l > v_l$ and $c_{si} < 1/v_l$ for $i=A$ or $B$,
where $v_l$ denotes the volume covered by a spacer.  It has been
approximated as\cite{danielsen2023phase} $v_l \sim (a \sqrt{l})^3$,
or, slightly more sophisticated\cite{book_doi_edwards}, as $v_l \sim
\frac{4}{3}\pi (a \sqrt{l})^3$. 
In our system, the parameters are
$a \sim 1.2 \sigma, \: v\sim2.2 \sigma^3$ and $l_A=5,\: l_B=10$. Hence, we have
$v\cdot l\sim 11 \sigma^3$ for RNA and $v\cdot l\sim 22 \sigma^3$ for trimers, which must
be compared to
$v_l \sim 19 \sigma^3$ or $\sim 81 \sigma^3$ for RNA 
and $v_l \sim 55 \sigma^3$ or $\sim 229 \sigma^3$ for trimer, depending on the
estimate.  Since $v \cdot l  < v_l$, excluded
volume effects can be considered small. On the other hand, the spacer
overlap concentration is $1/v_l\sim 0.05 \sigma^{-3}$ or $\sim
0.01 \sigma^{-3}$ for RNA, and  $1/v_l\sim 0.02 \sigma^{-3}$ or $\sim
0.004 \sigma^{-3}$ for trimers, depending on the estimate. The first estimate
corresponds to $c_{\text{RNA}} \sim 0.34 \sigma^{-3}$ and
$c_{\text{trimer}} \sim 0.14 \sigma^{-3}$, which is higher than most
concentrations considered in the present study. The second estimate
corresponds to $c_{\text{RNA}} < 0.08 \sigma^{-3}$ and
$c_{\text{trimer}} < 0.03 \sigma^{-3}$, which is comparable to our
values at intermediate RNA and trimer concentrations.}
\fsrev{Thus we
conclude that most of our systems are in the dilute regime, or on the
verge of being dilute, which explains the discrepancies between theory
and simulations at low concentrations. The mean-field theory also
assumes that the number of stickers on a chain is large $f \gg 1$,
which is clearly not true in the case of the trimers. However, this
factor seems to be less important.} At higher concentrations, the
theoretical predictions for the binding \xcrev{density} are in almost
quantitative agreement with the simulation results.

%We attribute this to correlation effects, which are prominent in dilute
%polymer systems. For example, the theory is obviously bound to break
%down at RNA concentrations much smaller than the overlap
%concentration, which is of order $c_{\text{RNA}}^* \propto
%N_{\text{RNA}}^{1-3 \nu} \sim 0.04 \sigma^{-3}$ (with the Flory
%scaling parameter $\nu = 0.588$). 

%\xcrev{According to the original Semenov–Rubinstein
%theory\cite{semenov1998thermoreversible}, the mean-field approach may
%lose validity when the excluded volume parameter $v$ is too large and
%$f-1\simeq  1$, because under such conditions the chain statistics
%deviate from Gaussian behavior and the spacer effect play a central
%role, causing the chains to swell and reducing the probability of
%sticker contacts (i.e., when $v/a^3>1/\sqrt{f/N}$). In our system, for
%both components, we have $v/a^3=2.2\sim 1/\sqrt{f/N}$
%($1/\sqrt{f/N}=2.55$ for RNA, and $2.77$ for trimer) and trimer chain
%contains too rare stickers. Consequently, the excluded volume effect
%decreases the binding density in low concentration region.  
%}

Next we consider the percolation transition. From the data shown in
Fig. \ref{fgr:P_c}, we determine the percolation threshold as the
point where \xc{$\langle P \rangle=0.5$}.  In other
studies\cite{choi2020generalized}, a fitting
function was utilized to estimate the percolation threshold.  We
tested that procedure for some cases and found the discrepancy between
this method and ours to be small. We also checked that the results are
not strongly affected by finite-size effects, indicating that the
simulation box is large enough\cite{wang2022phase}.

\begin{figure}[ht]
\includegraphics[width=7cm]{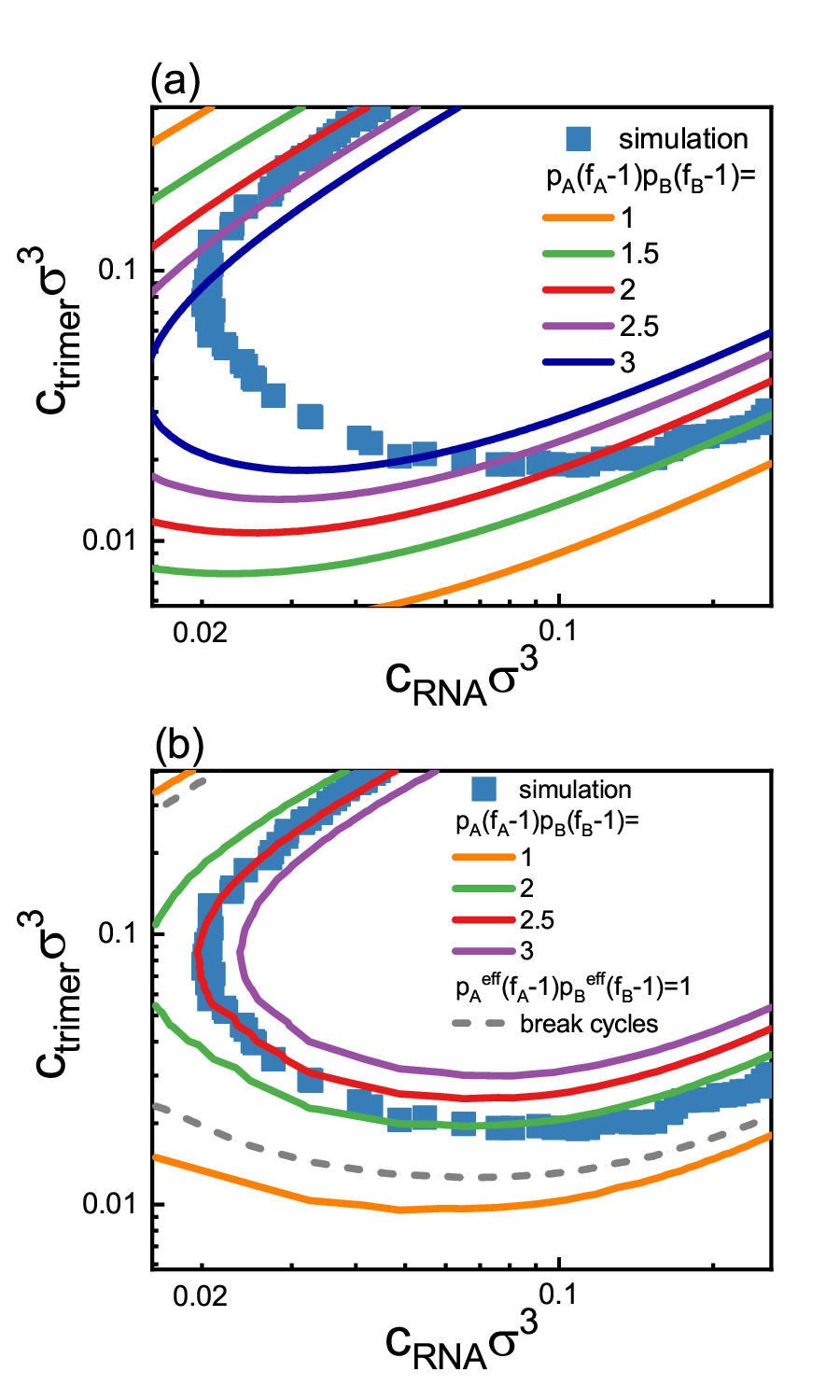}
\caption{Comparison of percolation threshold between the
simulation and theory. Blue square symbols are the results of the
simulation. Solid lines are the theoretical predictions based on Eq.\
(\ref{eq:P_c_theory}) for different values of the threshold value
$m_c$ as indicated and binding fractions $p_i$ taken from (a) theory
(Eq.\ (\ref{eq:p_i})) and (b) simulations. \xcrev{The gray dash line
shows the modification for $p_A(f_A-1)p_B(f_B-1)=1$. In this case, we
break all cycles (remain only one cycle for the spanning cluster) but
keep cluster connectivity in the entire system.} }
\label{fgr:P_c_theory}
\end{figure}

Fig. \ref{fgr:P_c_theory} shows the simulation data for the
percolation transition according to the definition above along with
the theoretical prediction of Eq.\ (\ref{eq:P_c_theory}).  This
equation relates the percolation threshold with the fractions
$p_{\text{RNA}}$ and $p_{\text{trimer}}$ of bound RNA and trimer
stickers, respectively, which are in turn calculated from Eq.\
(\ref{eq:p_i}). The theory predicts that the polymer network should
percolate if the average number of trimer-mediated connections between
RNA-chains, $m=18 \: p_{\text{RNA}} p_{\text{trimer}}$, exceeds the
threshold $m_p=1$. This prediction, shown as orange line in Fig.
\ref{fgr:P_c_theory}(a), greatly overestimates the concentration
region where percolation is observed. However, as we have argued in
Section II.B,
%\ref{sec:theory}
\xcrev{$m_p=1$ just gives lower bound for the onset of percolation.}
%percolating cluster and it does not consider any specific structures
%in the system.} 
\xcrev{The} true threshold value $m_p$ should be larger, \fs{for
example}, due to the existence of closed paths (cycles) in the
percolating cluster.
%\xcrev{and the excluded volume effect}. 
If we treat the threshold value as adjustable parameter and increase
it to $m_p=2$, the area of the percolated regime in the
$c_{\text{RNA}}$-$c_{\text{trimer}}$ plane shrinks, but the shape of
the percolated phase is still not captured well by the theory,
especially at low RNA concentrations.

On the other hand, we recall that this low concentration regime is
precisely the one where the theoretical predictions for the binding
numbers deviate from the simulation results (see Fig. \ref{fgr:theory}). Therefore, as a remedy, we also test the prediction
of Eq.\ (\ref{eq:P_c_theory}) if the values for bound sticker
fractions $p_{\text{RNA}}$ and $p_{\text{trimer}}$ are taken from
the simulations. The corresponding curves are shown as solid lines in
Fig. \ref{fgr:P_c_theory}(b), again together with the simulation
results for the percolation transition. One can see that the
theoretically predicted shape of the percolated region captures the
shape according to simulations much better. Assuming the threshold
value of $m$ to be at $m_p=1$, the size of the percolated region is
still too large, but we obtain reasonable agreement for $m_p \approx
2.5$. 

\xcrev{As we have argued previously, one possible reason for the
deviation -- even when using the binding numbers from the simulation
in Eq.\ (\ref{eq:P_c_theory}) -- is the presence of cycles in the
percolation cluster of the real system. To investigate the effects of
cycles, we have identified, in each configuration, the minimal bonds
necessary to maintain the connectivity of all clusters and removed all
the others, thus removing all cycles in the system. This was done
using the NetworkX Python package\cite{SciPyProceedings_11} and
Kruskal's algorithm\cite{kruskal1956shortest}. 
%use NetworkX to analyze the cycle information in the
In the gray dashed line in Fig. \ref{fgr:P_c_theory}(b), we compare
the results obtained for $m_p=1$ after removing the cycles (except for
one cycle for the spanning cluster) with the simulation results.
Here, the input parameter is the effective binding probability
$p_i^\mr{eff}=(N_p-N_\mr{break\ cycles}^\mr{max})/n_if_i$, where
$N_\mr{break\ cycles}^\mr{max}$ is the maximal number of bonds that
could be removed to break cycles without destroying clusters.  This
adjustment brought the theoretically predicted percolation line closer
to that obtained in the simulations, however, the theory still
substantially underestimates the percolation threshold.  Thus the
presence of cycles does not seem to be the primary cause of the
deviation.  }

To summarize, the mean-field theory can predict the binding number
fairly well, \xcrev{especially in dense solutions}, but the predicted
percolation threshold deviates strongly from the threshold located in
the simulations. This deviation can partly be attributed to
inaccuracies in the prediction for the binding number at small RNA
concentrations, and partly to the fact that the theory assumes a
tree-like graph structure of the percolating cluster at the transition
and neglects the possibility of cycles. \fsrev{However, these two
effects cannot fully explain the disagreement between theory and
simulations.} The theory underestimates the number of bonds needed to
create a percolating cluster. \fsrev{Interestingly,} good agreement
between theory and simulations can be obtained if this number is
treated as a single adjustable parameter.

%\xcrev{This
%deviation can primarily be attributed to the limitations of mean-field
%theory in good solvents, where the excluded volume effect plays a
%central role in reducing binding in the dilute region. Additionally,
%the theory provides only a lower bound for the transition and neglects
%the contribution of cycles, as not all bonds contribute to cluster
%expansion. } Therefore, \xcrev{in real systems, it is necessary to
%increase the threshold ($m>1$) to account for excluded volume effects
%under good solvent conditions.}

\subsection*{III.C Analysis of binding structures}

\begin{figure*}[ht]
\includegraphics[width=12cm]{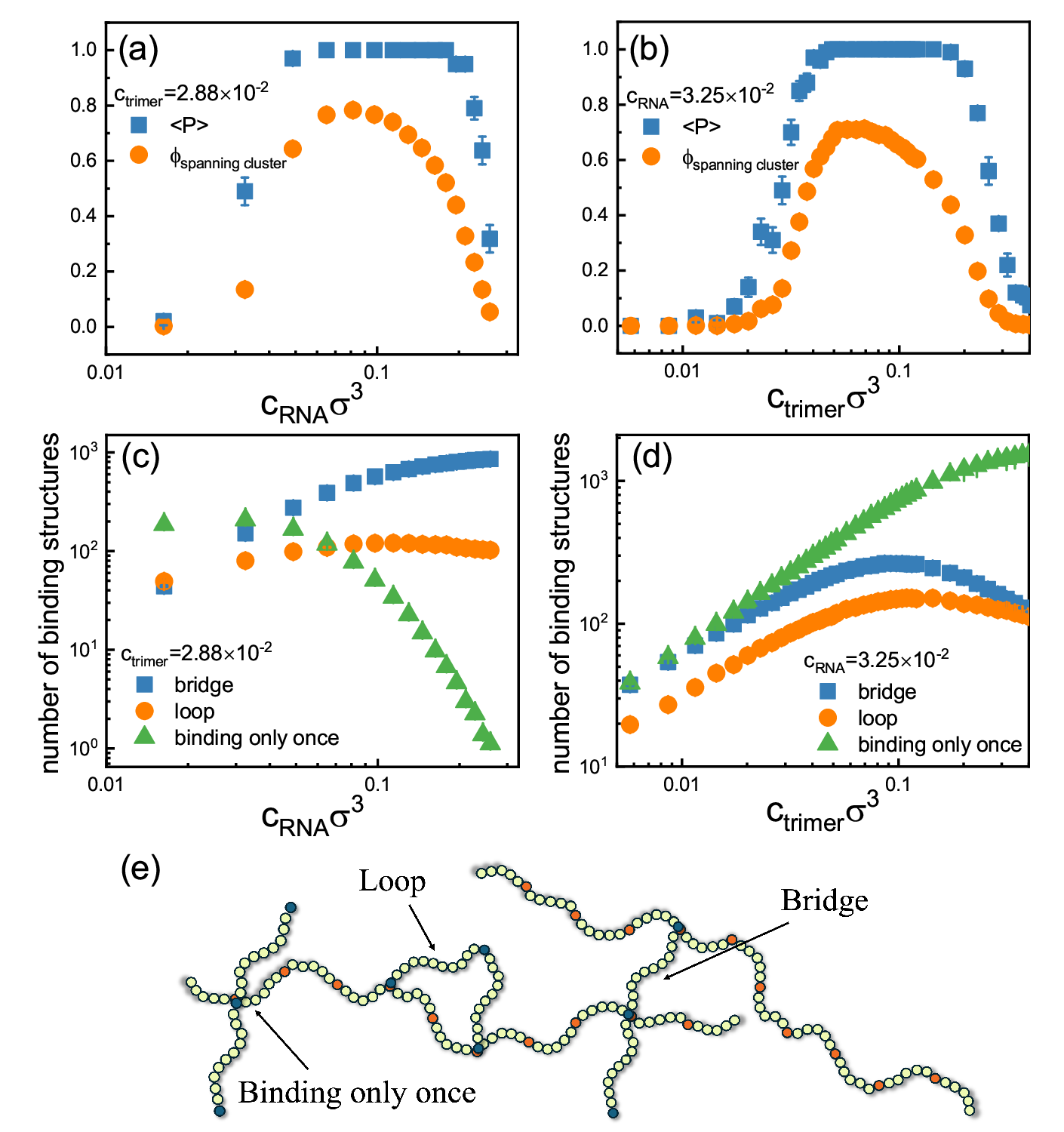}
\caption{(a,b) Order parameter \fsrev{$\langle P\rangle$ (blue
squares)} and size of spanning cluster $\phi$ \fsrev{(orange circles)}
in the case of fixed trimer \fsrev{bead} concentration
$c_\mr{trimer}=2.88\times 10^{-2}\sigma^{-3}$ and fixed RNA
\fsrev{bead} concentration $c_\mr{RNA}=3.25\times 10^{-2}\sigma^{-3}$.
(c,d) Corresponding number of binding structures. (e) Snapshot for
three binding structures as defined in the main text.}
\label{fgr:cluster_structure1}
\end{figure*}

Finally, we examine in more detail the molecular basis
of the reentrance behavior in the percolation, which is predicted by
theory\cite{danielsen2023phase} and observed in Fig. \ref{fgr:P_c}.
To this end, we define three binding structures: bridges, loops, and
''binding only once'' (see cartoon in Fig.
\ref{fgr:cluster_structure1}(e)). A bridge is formed when two stickers
on a trimer bind to two different RNA chains. A
loop occurs when two stickers on a trimer attach to the same RNA
chain. A trimer can be part of one or several bridges, or \xc{loops},
or both. If only one of its stickers is bound, it is classified as
''binding only once''.  Additionally, the relative mass of the
spanning cluster $\phi$ is defined as the ratio of the number of
particles in the spanning cluster to the total number of particles in
the system. We utilize the freud-analysis package\cite{freud2020} to
perform the cluster analysis.  Fig. \ref{fgr:cluster_structure1}
shows the details of the spanning cluster and the binding structures
for two cases.

Fig. \ref{fgr:cluster_structure1}(a) and (c) focus on the behavior
of the system at fixed trimer concentration as a function of RNA
concentration.  If the RNA concentration is low
($c_\mr{RNA}=1.63\times 10^{-2}\sigma^{-3}$), the number of bridges is
lower than that of loops, with the majority of trimers being attached
to only one RNA sticker. Thus, forming a spanning cluster is
challenging. As the RNA bead concentration increases beyond $c_\mr{RNA}>
0.06 \sigma^{-3}$, RNA and trimers connect to spanning clusters, where
bridges outnumber loops.  However, the total fraction of monomers in
the largest cluster never exceeds 80 percent.  Further adding RNA
reduces the spanning cluster's relative mass until
percolation breaks down at $c_\mr{RNA}=0.26\sigma^{-3}$. In this
reentrance region, the number of bridges and loops both reach
saturation, as all trimer binding sites become occupied.
The total number of specific bonds in the system saturates (cf.
Fig. \ref{fgr:theory}(a,c)). As a consequence, the {\em fraction}
$p_{\text{RNA}}$ of occupied RNA sticker sites subsequently decreases
with increasing RNA concentration, which, by virtue of Eq.\
(\ref{eq:P_c_theory}), leads to a \pb{reduction} of the average number
of trimer-mediated connections between RNA molecules and reduces the
percolation probability. At large RNA content, most RNA molecules only
bind to one trimer. Even though trimers still preferably form bridges
between \xc{RNA}, the number of bridges is not sufficient to create a
percolating network.

Owing to the structural similarity, the same reentrance behavior can
be observed at fixed RNA concentration, as shown in Fig.
\ref{fgr:cluster_structure1}(b) and (d). Increasing the trimer
concentration initially facilitates percolation, and the spanning
cluster size and bridge number reach a maximum at trimer concentration
$c_\mr{trimer}=6\times 10^{-2}\sigma^{-3}$. However, adding excess
trimers to the system subsequently leads to a situation where all
binding sites in RNA  are occupied by different trimers, so that the
number of trimers that bind only once keeps increasing,
while the numbers of bridges and loops decrease. 
The reorganization of binding structures eventually inhibits the
formation of spanning clusters.

\begin{figure}[ht]
\includegraphics[width=7cm]{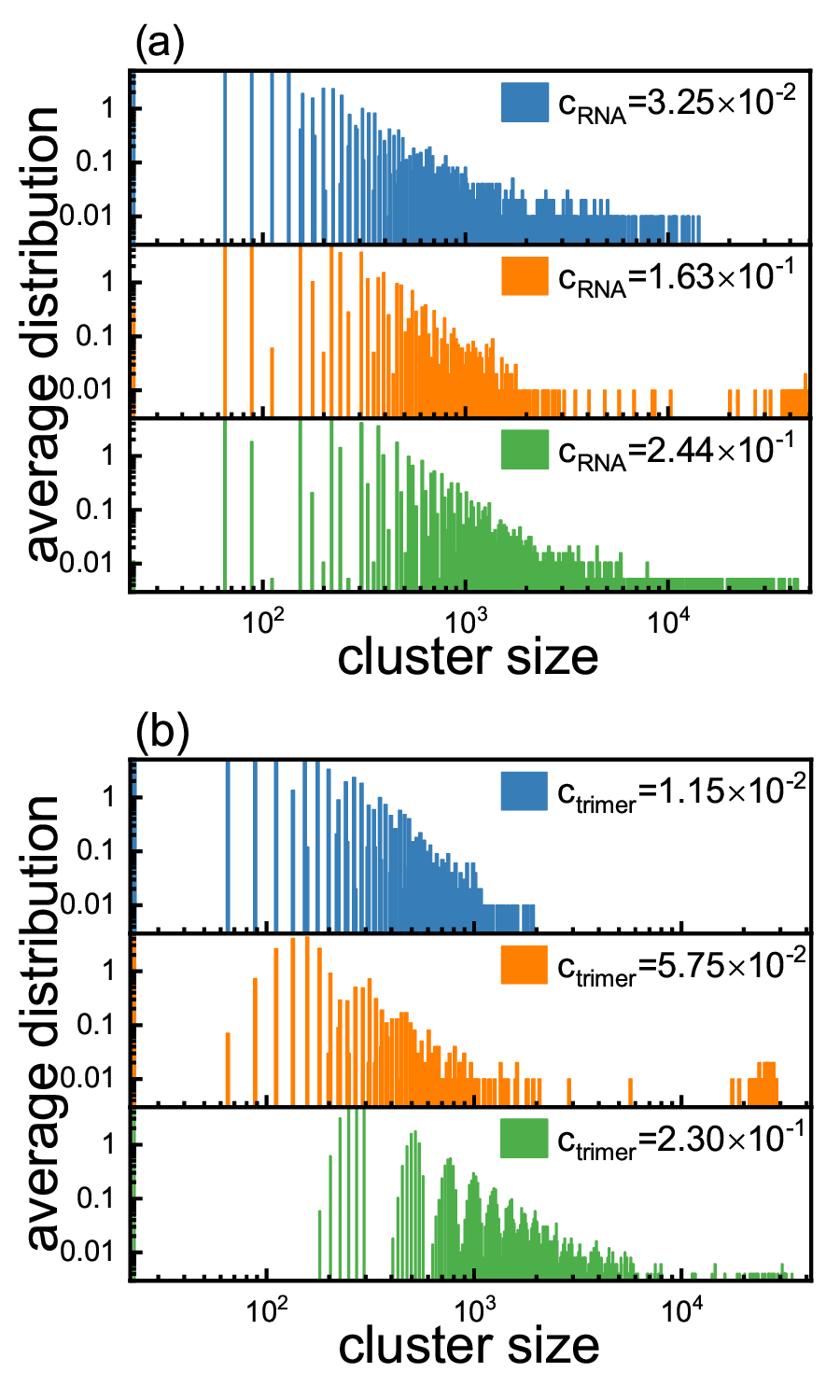}
\caption{Distribution of cluster sizes for the case fixed trimer
concentrations $c_\mr{trimer}=2.88\times 10^{-2}\sigma^{-3}$ (a) and
fixed RNA concentrations $c_\mr{RNA}=3.25\times 10^{-2}\sigma^{-3}$
(b), respectively.}
\label{fgr:cluster_structure2}
\end{figure}

To further analyze this phenomenon, we calculated the histogram of
cluster size distributions. \xcrev{Here, we average the histogram over
time}.
%to mitigate fluctuations in the cluster size distribution.  }
The results are shown in Fig. \ref{fgr:cluster_structure2}. At fixed
trimer concentration $c_\mr{RNA}=3.25\times 10^{-2}\sigma^{-3}$(top
panel in (a)), the system does not contain enough RNA molecules to
form a spanning cluster, hence the contribution of clusters containing
more than $2\times 10^{4}$ particles is not significant. In the
percolation region, clusters of intermediate size (containing $3\times
10^3$ to $2\times 10^4$ particles) decrease in number and merge into
larger clusters. As the RNA concentration continues to increase, the
system again predominantly forms smaller clusters, and at
$c_\mr{RNA}=2.44\times 10^{-1}\sigma^{-3}$(bottom panel in (a)), large
clusters become rare.  Fig. \ref{fgr:cluster_structure2}(b) shows
the evolution of cluster size distributions for fixed $c_{\text{RNA}}
= 3.25 \times 10^{-2} \sigma^{-3}$.  Here, adding more trimers leads to
the formation of more bridges that recruit RNA into the large
clusters, such that the largest cluster size exceeds $2\times 10^4$
particles(middle panel in (b)).  However, adding further trimers
results in a dominance of ''single binding'' instances, causing the
large clusters to break into several smaller clusters (bottom panel in
(b)). In summary, increasing the concentration of one component can
lead to reentrant behavior in the phase diagram. The size of the
largest cluster decreases, and eventually, percolation is no longer
observed. If the trimers are in excess, this is associated with a
reorganization of binding structures as described above, such that the
bridges lose their dominant role. If RNA is in excess, the bridges
remain dominant, but the number of bridges per RNA molecule is too
small to sustain percolation. \fsrev{The results suggest that there
exist two cluster populations in the percolation region: 
one population with a size distribution that decreases roughly 
exponentially as a function of cluster size, and one featuring
a distinct peak at a large, ''macroscopic'' cluster size. 
Outside the percolation region, the latter peak disappears and 
all cluster sizes are exponentially distributed. Therefore, the 
existence of this peak might be a good alternative criterion 
that could be used to identify percolation transitions in real finite 
systems, where the spanning cluster criterion cannot be applied.}

\section*{IV Summary}

The main purpose of the present study was to investigate the impact of
specific binding on the phase behavior of heteroassociative polymer
solutions, such as solutions of RNA and proteins containing domains
that can selectively bind to RNA. To this end, we have utilized
molecular dynamics simulation to study the percolation transition
without phase separation in RNA-protein mixtures, and compared the
results with the mean-field theory of associative polymers.  Our main
results can be summarized as follows:

\begin{enumerate}[(1)]
\item Percolation can be observed in heteroassociative mixtures if
the concentration of both components is sufficiently high. In the
systems considered in the present work, the percolation was not
accompanied by liquid-liquid phase separation.

\item The percolation phase diagram features double reentrance
behavior with increasing each concentration of either the RNA or the
protein component. Protein mediated bridge formation is identified as
being crucial for cluster formation and the percolation transition.
If one of the two components is in excess of the other, percolation is
suppressed: If proteins are in excess, bridges do not form. If RNA is
in excess, the number of bridges per RNA molecule is not sufficient.
In both cases, this inhibits percolation.

\item The mean field theory for associative polymer can be extended to
two asymmetric components and can be used to predict the behavior of
the RNA-protein system. The theory successfully captures the evolution
of binding numbers as a function of the concentrations of the two
components, \xcrev{especially at high concentrations,} and explains
the stability of the homogeneous system. 
%\xcrev{even if the theory is not strictly valid under good solvent
%condition}.  
It shows how spacers between binding sites on the RNA and the proteins
play a crucial role in modulating the competition between specific and
unspecific monomer interactions, thereby enhancing the solubility of
each component and preventing phase separation in our system. However,
the theory fails to accurately predict the percolation threshold. This
discrepancy can be attributed in part to small
deviations between the average number of
specific bonds as obtained from simulations and theory \fsrev{due to
the fact that the concentrations of the polymers are in a regime where
the conditions for the validity of the mean-field theory are not
fulfilled. Another factor is the existence of} closed, paths in the
percolating cluster at the percolation threshold, \fsrev{which is
neglected by the theory. However, our detailed analysis shows that
these two factors cannot fully account for the discrepancy between
theory and simulations. Interestingly,} good agreement between theory
and simulations can be obtained by introducing a single heuristic fit
parameter \fsrev{into the criterion for the percolation threshold.}

\end{enumerate}

Previous studies have often focused on the phase separation behavior of
biomolecular systems.  The present paper enhances the understanding of
the mechanisms underlying the formation of percolating networks
involving RNA and proteins. For proteins, many factors control the
phase behaviors, such as the sequence of IDRs and RBDs, the solubility
of IDRs, the binding specifics of RBDs, etc. In our study, we only
discussed the effect of specific binding between RBDs of proteins and
RNA, and we did this at a very generic level.  We did not consider the
impact of nonspecific binding, which may lead to the formation
metastable, long-living droplets in sticker-spacer models for proteins
according to the literature\cite{ranganathan2020dynamic}.  Moreover,
we used a simple simulation model of fully flexible chains and did not
consider possible effects of chain rigidity, e.g., of RNA.  Another
interesting topic for future studies is sequence-dependent gelation or
phase separation. For example, Jain and Vale found that RNA with
longer repeats(GGGGCC) formed an interconnected mesh-like network,
while RNA with different repeats(CCCCGG) was
soluble\cite{jain2017rna}.  Seim et al.  reported that for a fungal
RNP protein, Whi3, the formation of transient alpha-helical structures
also modulates phase separation by promoting the assembly of dilute
phase oligomers\cite{seim2022dilute}. According to this literature,
the sequence of the special binding motif in RNA or proteins also
emerges as a crucial determinant in determining the phase transitions
in the system.

We have shown that the Semenov-Rubinstein theory provides reasonable
predictions of the behavior of associative polymers in the semidilute
regime, but it cannot capture correlations in the dilute regime, where
stickers are no longer evenly distributed in the system.  Moreover, it
neglects the conformational entropy of chains. Recently, Rovigatti et
al. studied the role of entropy in the phase behavior of systems of
associating single-chain nanoparticles\cite{rovigatti2022designing}
and showed that the configurational and combinatorial binding entropy
can modulate the phase behavior.

In summary, our approach provides basic insights into the unique phase
behaviors of RNA-protein systems driven by specific binding, setting
the stage for more detailed investigations into the factors governing
these phenomena including their theoretical description.


%\begin{acknowledgement}
\begin{acknowledgments}

This project was funded by SFB 1551 Project No. 464588647 of the DFG
(Deutsche Forschungsgemeinschaft). The authors gratefully acknowledge
the computing time provided to them on the high-performance computer
Mogon2 and Mogon NHR South-West. 

\end{acknowledgments}
%\end{acknowledgement}

% \begin{suppinfo}

% A listing of the contents of each file supplied as Supporting Information
% should be included. For instructions on what should be included in the
% Supporting Information as well as how to prepare this material for
% publications, refer to the journal's Instructions for Authors.

% The following files are available free of charge.
% \begin{itemize}
%   \item Filename: brief description
%   \item Filename: brief description
% \end{itemize}

% \end{suppinfo}

\bibliography{main}

\end{document}